\documentclass[fleqn,usenatbib]{mnras}

\usepackage{newtxtext,newtxmath}
\usepackage[T1]{fontenc}
\usepackage{ae,aecompl}

\usepackage{soul}

\usepackage{graphicx}	% Including figure files
\usepackage{amsmath}	% Advanced maths commands
\usepackage{amssymb}	% Extra maths symbols
\usepackage{tcolorbox}
\title[Haumea's ring]{Dynamics of Haumea's dust ring}

\author[T. Kov\'acs and Zs. Reg\'aly]{
T. Kov\'acs$^{1}$\thanks{E-mail: tkovacs@general.elte.hu (TK)}
 and Zs. Reg\'aly$^{2}$
\\
$^{1}$Institute of Theoretical Physics, E\"otv\"os University, 1117, Budapest, P\'azm\'any P. s. 1A, Hungary\\
$^{2}$Konkoly Observatory, Research Centre for Astronomy and Earth Sciences, \\\,\,\,\,\,Hungarian Academy of Sciences, 1121, Budapest, Konkoly Thege Mikl\'os \'ut 15-17, Hungary\\
}

\date{Accepted XXX. Received YYY; in original form ZZZ}

\pubyear{2018}

\begin{document}
\label{firstpage}
\pagerange{\pageref{firstpage}--\pageref{lastpage}}
\maketitle

\begin{abstract}
The particle dynamics of the recently observed ring around dwarf planet Haumea is numerically investigated. The point mass gravitational force, a second degree and order gravity field, and the solar radiation pressure as the main perturbations are considered. The quasi-stationary state of the ring varies for different micron-sized grains and depends also on the spin-orbit resonances between the rotation rate of the main body and the orbital period of the dust particles. The simulations confirm the variable radial width of the ring observed during the transit ingress and egress. Results show that the micron sized grains, initially on circular orbits, become eccentric and form an apse-aligned ring at the observed radial distance near to the 3:1 spin-orbit resonance. It is also demonstrated that this coincidence is only apparent and independent of 3:1 resonance.   
\end{abstract}


%%%%%%%%%%%%%%%%%%%%%%%%%%%%%%%%%%%%%%%%%%%%%%%%%%

%%%%%%%%%%%%%%%%% BODY OF PAPER %%%%%%%%%%%%%%%%%%

\section{Introduction}

Observations performed in last two decades confirm \citep{Braga2014} and propose \citep{Ortiz2015}\footnote{The interpretation of Chiron's ring is still an open issue.} ring systems around Centaurs as well as around small bodies from the Trans-Neptunian region \citep{Ortiz2017}. So far the ring's origin \citep{Araujo2016,Michikoshi2017,Wood2017}, dynamical evolution \citep{Pan2016}, composition and other physical parameters such as the grain size, mass, colour and albedo \citep{Stern2009} have been investigated. Recently, a circular ring around Haumea has been discovered by \citet{Ortiz2017}. They claim that the ca. 70km wide ring around the triaxial object is located near to the 3:1 spin-orbit resonance related to the planet's rotation frequency and the ring particles' mean motion around it. 

In the solar system the particles' orbit around a uniformly rotating triaxial object, is affected by various perturbation forces such as oblateness of planet, radiation pressure, tidal forces, or magnetic field \citep{Hamilton1996}. Consequently, the circumplanetary dust grains can have a reach dynamics fairly different from the Keplerian motion even in two-body case \citep{Scheeres1994,Feng2017}. The fate of the particles as well as the shape and structure of the ring depends strongly on the relative strength of these perturbations. 

The goal of this paper is to present a simple dynamical model of the proposed ring system of Haumea based on \citet{Hu2004} and investigate the resulting structure under the dominant perturbations. Our results are consistent with the pioneering observations presented in \citet{Ortiz2017} and shed light on the dynamics behind the observed structure of the ring. First, we describe the physical model and introduce the numerical setup. The ring morphology based on numerical calculations is presented in next section followed by our conclusions and outlook to the possible extension of the model for the future studies. 

\section{Model, methods, and simulations}

\subsection{Basic assumptions}

\begin{figure}
\centering
	\includegraphics[width=1.0\columnwidth]{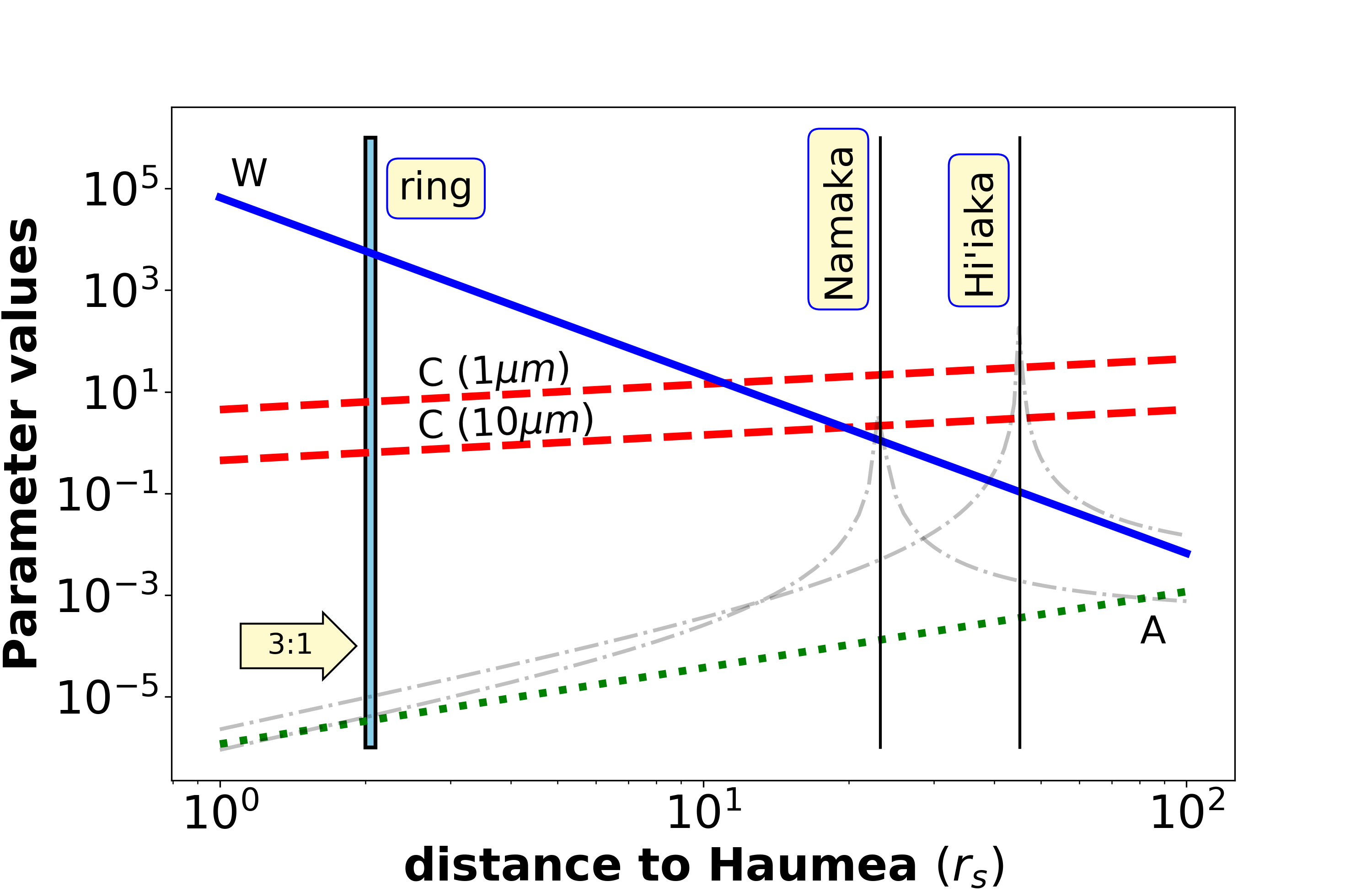}
\caption{Comparison of dimensionless parameters that affect grain dynamics as a function of the orbital radius. The oblateness (W) and the solar radiation pressure (C) are taken into account in the simulations while the solar tidal force (A) and the gravitation of the moons  are neglected (gray dashed-dotted lines). The position of the proposed ring and the moons' semimajor axes are also indicated. Length unit $r_s\approx$ 1103km is defined in Eq.~(\ref{eq:units}).} %r=1.75, t=1000
    \label{fig:1} 
\end{figure}

\begin{figure*}
\centering
	\includegraphics[width=2.0\columnwidth]{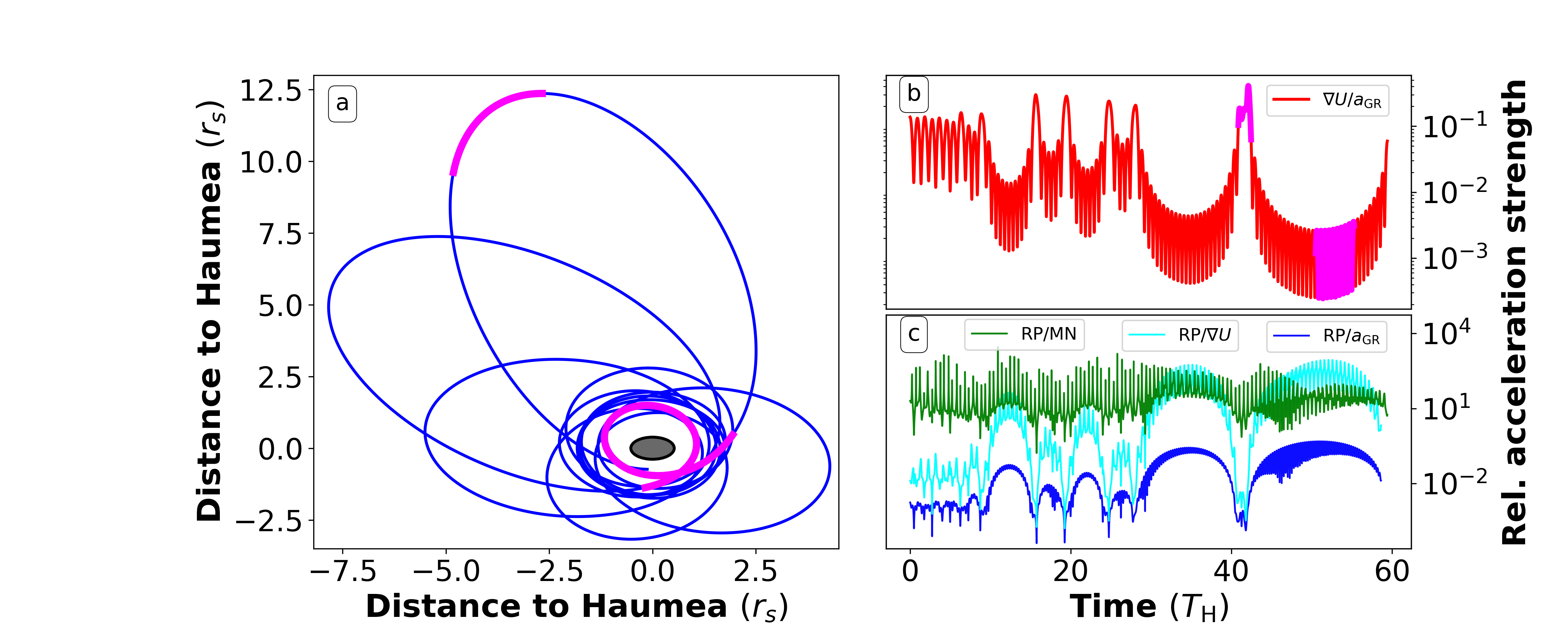}
\caption{Relative strength of different perturbations. (a) Orbit evolution projected into $x-y$ plane. Initial conditions of circular orbit: $x
_0$=1.75, $y_0$=0. (b) Comparison of $a_{\mathrm{GR}}$ and the acceleration $\nabla U.$ Peaks correspond to close encounters, one of them (t$\approx$ 40) highlighted in panel (a) (purple thick segment). For completeness a farther part of the orbit is also marked (t$\approx$ 53). (c) Additional perturbations, radiation pressure (RP) relative to $a_{\mathrm{GR}},$ and $\nabla U,$ and the moons' gravitational attraction (MN). The time is measured in Haumea's rotation period, $T_{\mathrm{H}}.$}
    \label{fig:2} 
\end{figure*}

In order to study the ring dynamics around Haumea first we describe the perturbations to the ring particles following \cite{Hamilton1996}. We omit, however, the Lorentz parameter, the ratio between the Lorentz force and the planet's gravity, by assuming that grains do not have electric charge. Furthermore, the solar tidal force, described by the parameter $A=0.75n_{\odot}/n$ (with notations in Eq.~(\ref{eq:obl})), can also be neglected close to the planet where the ring is formed, see Fig.~\ref{fig:1}. We take into account the gravitational potential of a uniformly rotating body and the stellar radiation force. The former is determined by the Haumea's shape that can be considered as a triaxial ellipsoid with semi-axes $a=\text{1161}\pm\text{30 km}$ $b=\text{852}\pm\text{4 km}$ $c=\text{513}\pm\text{16 km}$ \citep{Ortiz2017}. The later is determined by the ratio of the stellar radiation and gravitational forces, see Equation~\ref{eq:rad}. The solar radiation causes the Poynting--Robertson drag (PR) and the radiation pressure (RP). The PR drag is excluded from our analysis as it has considerable effect only for longer time scales (> 10$^5$ yr) that is out of scope of the current work.

The oblateness parameter is
\begin{equation}
W=-\frac{3}{2}C_{20}\left(\frac{R}{a}\right)^{2}\frac{n}{n_{\odot}},
\label{eq:obl}
\end{equation}
where $C_{20}$ and $R$ denote the second zonal harmonic coefficients and the effective radius of the planet, respectively, while $n$ is the mean motion of a given particle orbiting the planet, $n_{\odot}$ is the mean motion of the planet around the sun. For simplicity and loss of generality at this point we consider the planet as an oblate spheroid. The magnitude of $W$ is in the same magnitude as with triaxial shape.

The quantitative form of the radiation parameter $C$ reads as follows
\begin{equation}
C=\frac{9}{8}\frac{n}{n_{\odot}}Q_{\mathrm{PR}}\frac{F_{\odot}r^2}{GMc\rho s}.
\label{eq:rad}
\end{equation}
The radiation pressure efficiency parameter $Q_{\mathrm{PR}}$ is considered to be 1 that holds for ideal absorbing particles \citep{Burns1979}. In general, for different real materials (silicates, water ice, iron, magnetite, etc.) the value of $Q_{\mathrm{PR}}$ depends on the optical properties. $F_{\odot}$ is the solar radiation flux density at the heliocentric distance of Haumea. Furthermore, $r$ denotes the absolute value of the position vector of the particle, $G$ is the gravitational constant, $M$ is Haumea's mass, $c$ the speed of light. Parameters $\rho$ and $s$ refer to the particle's bulk density and radius, respectively. 

It is evident that the contribution of radiation pressure is proportional to the particle's distance squared and inversely proportional to the grain size. As a result, the smaller the particle, the larger the effect of RP. Moreover, for a given grain size RP is more dominant at larger distances to the Haumea. For a complete picture on how the parameters $C,\;W,$ and $A$ depend on the distance and grain size see Figure~\ref{fig:1} in \citep{Hamilton1996} or \citep{dosSantos2013}.

Figure~\ref{fig:2} shows the time dependence of the acceleration components and their relative strengths for a segment of a particular orbit. The left part (a) presents the orbit of a particle starting on circular orbit in the equatorial plane of the planet. Two segments are marked at two different positions in order to show the evidence of the contribution of the oblateness versus the point-mass gravitational effect. Panel (b) shows $|\nabla U|$ determined from Eq.~(\ref{eq:pot}) compared to Haumea's gravitational attraction, $a_{\mathrm{GR}},$ to the test particle as if it were a point mass . It is clear that $a_\mathrm{GR}$ dominates the dynamics. However, the component $\nabla U$ provides a significant amount up to a factor of 0.5 at close encounters, for example at $t\approx$ 40. 

Two other components of acceleration are also considered in panel (c): radiation pressure (RP), and the gravitational influence of the two moons of Haumea, Hi'iaka and Namaka (MN). As one can see, the RP does not play a significant role to the dynamics when the particle is close to the main body (RP/$a_{\mathrm{GR}}$). In contrast, far from Haumea, RP becomes significant (20-30\% of $a_{\mathrm{GR}}$). %the leading role. 
The effect of the moons (RP/MN), i.e. the radiation pressure vs. their gravity, can be safely excluded. Calculations of this later comparison were done by assuming the physical and orbital parameters of the moons given in Table~\ref{tab:moons} according to \cite{Ragozzine2009}.

\begin{table}
	\centering
	\caption{Physical parameters of the moons of the dwarf planet Haumea.}
	\label{tab:moons}
	\begin{tabular}{lccc} % four columns, alignment for each
		\hline
		Name & mass & semimajor & eccentricity\\
        & (x10$^{18}$kg) & axis (km) & \\
		\hline
		Namaka & 1.79$\pm$1.48 & 25657$\pm$91 & 0.249$\pm$0.0015\\
		Hi'iaka & 17.9$\pm$1.1 & 49880$\pm$198 & 0.0515$\pm$0.0078\\
		\hline
	\end{tabular}
\end{table}

\subsection{Equations of motion and numerical setup}

To model the ring dynamics we use the dimensionless equations of motion in a frame that co-rotates with the central body whose principal moments of inertia are $I_{xx}\leq I_{yy}\leq I_{zz}$ with axes $x,y,\text{ and } z$ \citep{Hu2004}. Introducing the characteristic length scale $r_s\approx$ 1103 km, being the synchronous orbit where the point mass gravitational attraction of the dwarf planet equals to the centripetal acceleration
\begin{equation}
r_s=(\mu/\omega_{\mathrm{H}}^2)^{1/3},
\label{eq:units}
\end{equation}
where $\omega_\mathrm{H}$ is the angular speed of the rotation of Haumea. The characteristic time-scale is set to be the rotational period of Haumea $T_\mathrm{H}=2\pi$, in which case $\mu=G M=$1.

\begin{figure*}
	\centering
	\includegraphics[width=2.0\columnwidth]{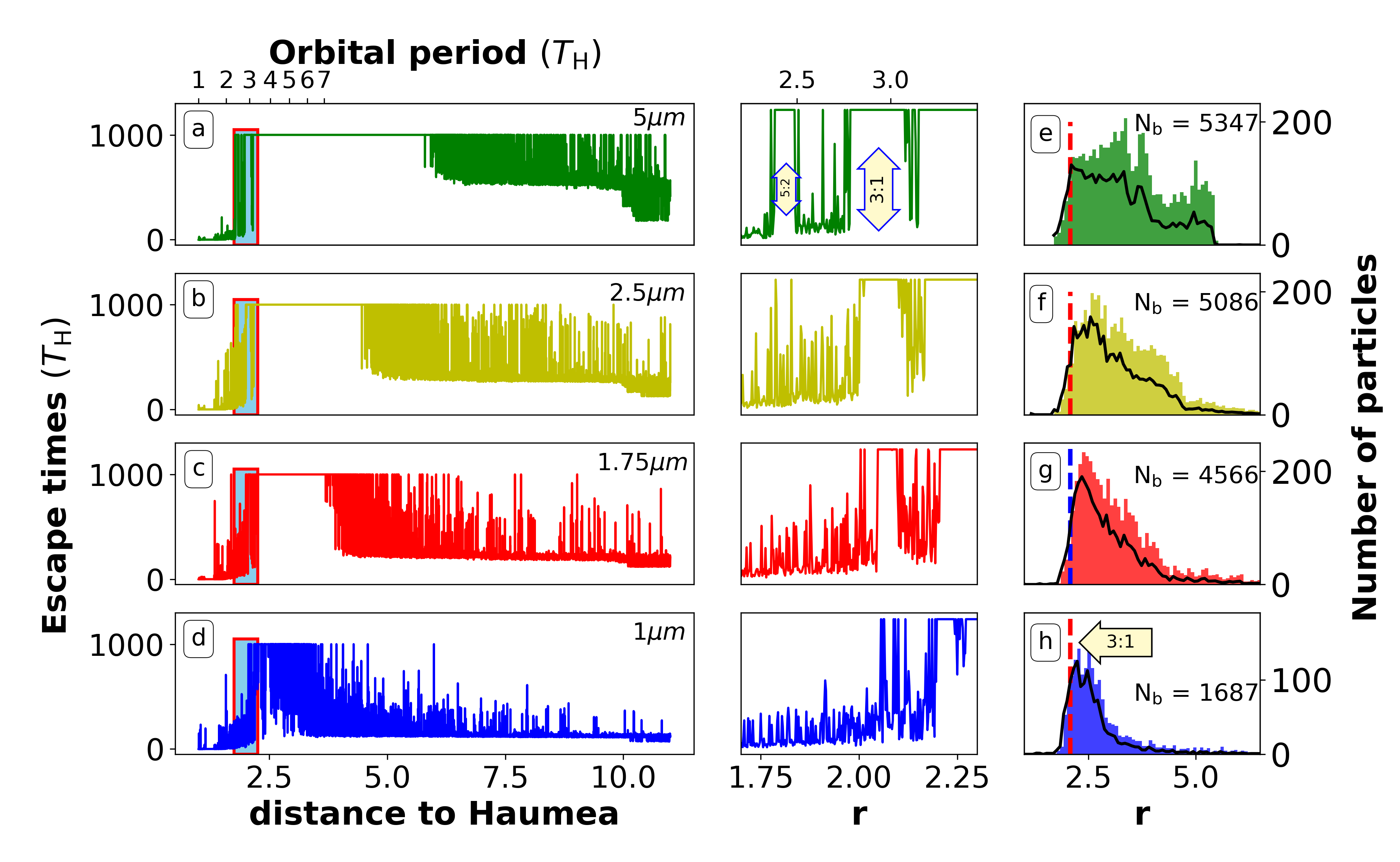}
    \caption{Left: Particles' escape time as a function of distance along the $x$-axis (i.e. $y$=0), (1<$r$<11). Middle: Zoom in of regions bounded by rectangles shown in (a)-(d). Right: Radial distribution of the survived particles in various sizes: (e) $5\,\mu$m, (f) $2.5\,\mu$m, (g) $1.75\,\mu m$, and (h) $1\,\mu$m. Solid black curve superimposed to the histograms indicates the surface density. The number of survived particles, out of initial 10000, are indicated with $N_\mathrm{b}$. The gray ellipse represents Haumea to scale. Red (blue in panel (g)) dashed lines mark the 3:1 spin-orbit resonance.} % 1687,4566, 5086, 5347 // 314, 1029, 1489, 2396
    \label{fig:3}
\end{figure*}

The potential of Haumea is approximated by a uniformly rotating second order and degree gravitational field with coefficients $C_{20} \leq 0$ (zonal potential term) and $C_{22} \geq 0$ (equatorial ellipticity potential term) which are directly linked to the principal moments of inertia through the semi-axes \citep{Balmino1994} as
\begin{equation}
\begin{split}
C_{20} &= -0.5(2I_{zz}-I_{xx}-I_{yy})/r_s^2 = -0.1274,\\
C_{22} &= 0.25(I_{yy}-I_{xx})/r_s^2 = 0.0256,
\end{split}
\end{equation}
where the normalization factor, the Haumea's mass, is used. Following \citet{Hu2004} the normalized equations of motion read
\begin{equation}
\begin{split}
\ddot{x}-2\dot{y}&=&x-\frac{x}{r^3}+\frac{\partial U}{\partial x}+F_{x,\mathrm{rp}},\\
\ddot{y}+2\dot{x}&=&y-\frac{y}{r^3}+\frac{\partial U}{\partial y}+F_{y,\mathrm{rp}},\\
\ddot{z}&=&-\frac{z}{r^3}+\frac{\partial U}{\partial z}+F_{z,\mathrm{rp}}\\
\end{split}
\label{eq:eom}
\end{equation}
with normalized force potential
\begin{equation}
U=-\frac{C_{20}(x^2+y^2-2z^2)}{2r^5}+\frac{3C_{22}(x^2-y^2)}{r^5},
\label{eq:pot}
\end{equation}
where $r=\sqrt{x^2+y^2+z^2}$ is the particle's position vector.

The solar radiation pressure, $\mathbf{F}_{\mathrm{rp}}$, is also taken into account. Since a co-rotating frame is used, $\mathbf{F}_{\mathrm{rp}}$ is time-dependent
\begin{equation}
\mathbf{F}_{\mathrm{rp}}=m_p\dot{\mathbf{v}}=\frac{F_{\odot}A Q_{\mathrm{pr}}}{c}\mathbf{\hat{r}}(t),
\end{equation}
where $m_p=(4\pi/3)\rho s^3$ is the particles mass orbiting with velocity $\mathbf{v},$ $A$ is the particle's cross section, and $\mathbf{\hat{r}}(t)$ is the unit vector pointing to the direction of the radiation. The particles are assumed to be spherical with uniform density equal to 1 g/cm$^3.$ For transparency we give the explicit form of $\mathbf{\hat{r}}(t)$
\begin{equation*}
\mathbf{\hat{r}}(t)=(\cos -\omega_{\mathrm{H}}t,\sin -\omega_{\mathrm{H}}t,\cos n_{\mathrm{H}}t\sin i),
\end{equation*}
$\omega_{\mathrm{H}}$ is Haumea's rotational frequency, $n_{\mathrm{H}}$ is the mean motion about the Sun, and $i$ is the obliquity of the dwarf planet to its orbital plane.

The gravitational effect of the Sun is not included in this model because it is 5 order of magnitude weaker than that of Haumea at the distance of the ring. However, Haumea's orbit is assumed to be circular with a period of 284 years which is included in the time dependence of  the RP. Since the obliquity of the dwarf planet is about 14 degrees this implies a long-term modulation in z-component of the radiation pressure. In other words, the initial position of the Sun in our calculations is the negative $x$ direction and Haumea's obliquity stays in this direction during the whole integration.

The effect of planetary shadowing (which blocks the stellar radiation impinging on the ring) is neglected since the ring is currently not shadowed because it is in the equatorial plane of the planet which is tilted with 13.8$^\circ\;\pm$ 0.5$^\circ$ to the orbital plane \citep{Ortiz2017}. Furthermore, even for longer integration, only a small fraction of the ring will be hided during Haumea's journey around the sun.

We use a 4th order Runge-Kutta numerical scheme with a fixed step size that keeps the relative energy error below 10$^{-8}.$ Equations~(\ref{eq:eom}) are integrated for different times in order to cover different dynamical regimes. That is, the simulations are performed for $T$=10$^3$, 2.5$\times$10$^4$, and 3.5$\times$10$^5$ rotation periods, yielding ca. 1, 10, 1000 years, respectively.%, which is in the order of typical grain lifetimes. 
The calculations also have been done for grain sizes, 1, 1.75, 2.5, and 5 $\mu$m. The dynamical analysis involves two different setup of initial conditions. In the one hand, when studying the fine resonant structure, particles are placed equidistantly along a line (x axis) in radial distance 1<$r$<11. On the other hand, an initial ring is defined, $\theta\in[0;2\pi]$ and $r\in[0;5],$ in order to follow the particle dynamics in general. In both cases grains are initially on circular Keplerian orbits in the equatorial plane of Haumea.

\section{Ring characteristics}

Our first analysis explores the lifetime of the particles around the dwarf planet. The grain dynamics yields three different end-states for the particles:
(i) \textit{Escape}: When the semi-major axis of a particle becomes larger than 25$r_s$ ($\approx$ 25575 km) it is taken to be ejected. Ejections occurs due the close encounters with Haumea when the particles are perturbed to hyperbolic orbits. (ii) \textit{Collision}: In our model a collision occurs when the particle approaches the dwarf planet closer than its volume-equivalent radius, 816 km (0.74$r_s$) \citep{Ortiz2017}. (iii) \textit{Bounded motion}: When a particle, until the end of the integration, neither collides nor escapes from the system its motion is considered to be bounded.

Figure~\ref{fig:2} shows the escape times as a function of $r$ for four different grain sizes, 5, 2.5, 1.75, and 1 $\mu$m, see panels a, b, c, and d, respectively. The plots have been generated as follows: initially 5000 particles are placed equidistantly in radial direction along the x-axis in the interval $1<r<11.$ Their motion is followed until a collision or escape occurs or the integration time is over ($T$=1000 rotations of Haumea).

\begin{figure}
	\centering
	\includegraphics[width=1.0\columnwidth]{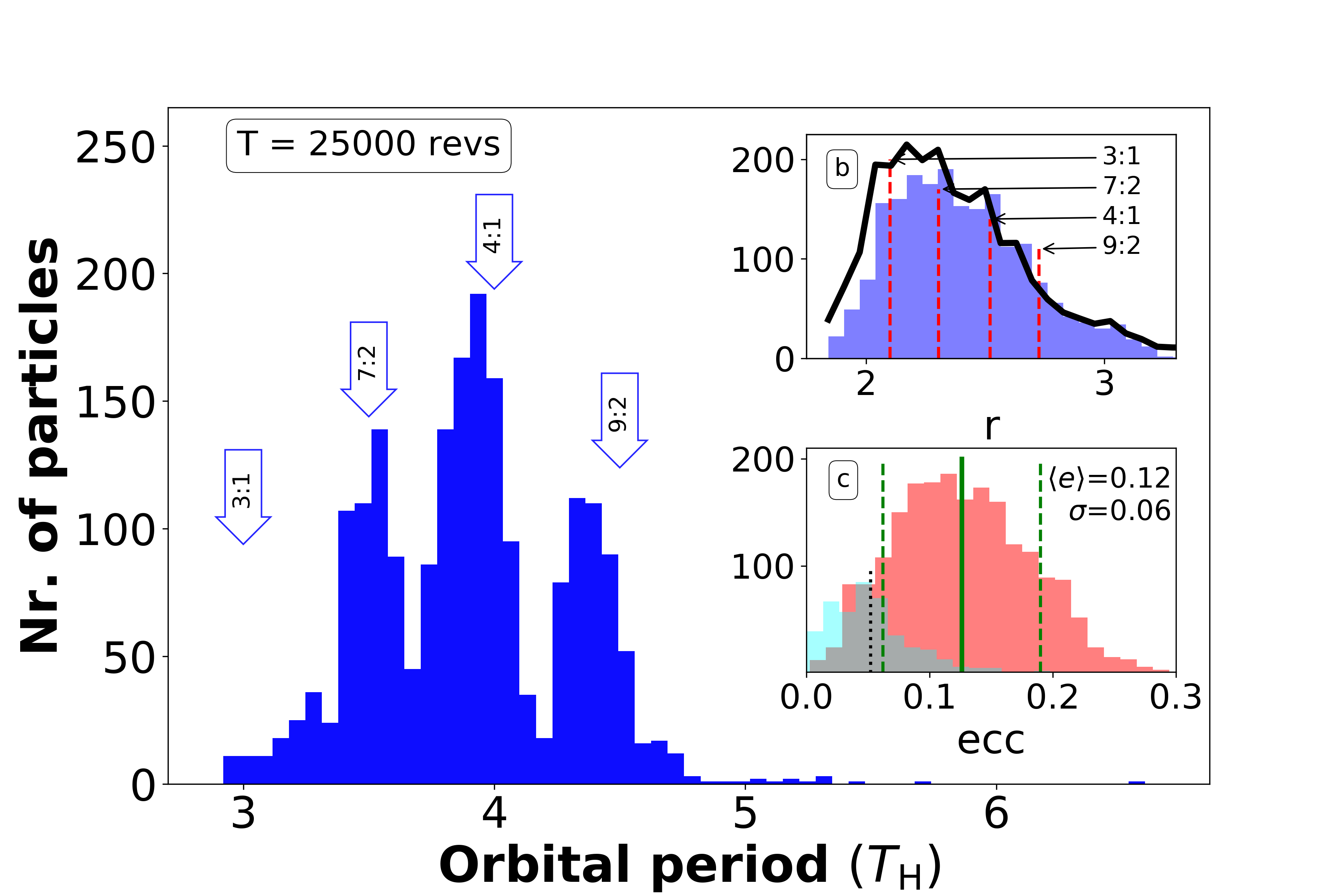}
    \caption{Number of the survived particles (1$\mu m$) vs. their orbital period measured in Haumea's rotation time. Panel (b): Radial distance distribution and the surface density (black). Panel (c): the eccentricity distribution of the bounded grains at the end of the integration (red). The blue histogram shows the distribution when no RP is present.} %2028 out of 60000
    \label{fig:4}
\end{figure}

One can observe longer and shorter plateaus extending between $2<r<5.5$ for different grain sizes (except for 1$\mu m$ in panel d where no plateau is found in this resolution). These plateaus correspond to bounded orbits, i.e. particles survive until the end of the integration. The larger the grain, the longer the plateau.

The spiky structure of the post-plateau regions indicates that the particles from this region escape very soon. Quantifying the rapid decay, it turns out to be exponential. This phenomenon is well-known in open dynamical systems where the dynamics is governed by unstable Lyapunov orbits \citep{Contopoulos2002,Lai2011}. The characteristic lifetime of a particle obtained from the decay rate is, for example, ~1050 rotations (ca. 0.5 years) in case of 1$\mu$m particles (from the domain $2 \leq r\leq 4$ in Fig.~\ref{fig:2} panel d). This rapid depletion of micron sized particles has also been observed by \citet{dosSantos2013} in case of Pluto's moons.
Moreover, the effect of the radiation pressure increases with $r.$ The sudden drop in survival time beyond r$\approx$10 yields the C$\approx$1, where the gravitational force of Haumea and the radiation pressure becomes commensurable, see Eq.~(\ref{eq:rad}). 

By magnifying the inner edge of the plateau for each grain size it is appreciable that the dynamics becomes very reach closer to Haumea (Figure~\ref{fig:3} middle panels).
On the one hand, for larger particles several spin-orbit resonances emerge due to the perturbation caused by the oblate central body. The shorter plateau around $r\approx$1.8 corresponds to the 5:2 resonance while the longer one at $r\approx$2.1 is the 3:1. Thus, particles starting near to spin-orbit resonances remain a member of the ring, while the others, starting farther from the resonances, escape or collide to Haumea.

On the other hand, considering smaller grains (panels b, c, d) first the 5:2 then the 3:1 resonance disappears. This can be explained by a growing contribution of the RP to the particle's dynamics. As a result, not shown here, bounded orbits can only originate beyond the 3:1 spin-orbit resonance for 1$\mu $m grain size. That is, only short plateaus farther from Haumea survive, and protect the particles from escape or collision, see the details below. 

Now let us concentrate to the bounded orbits. In order to have a reliable statistics, 10$^4$ particles are placed uniformly in a ring (1$\leq r\leq $5, 0$\leq \phi \leq 2\pi$) around the planet.

Panel h in Figure~\ref{fig:3} shows the histogram of the final positions (azimuthally averaged) of 1 $\mu$m size grains. As one can see the 1687 survivors produce a sharp peak slightly beyond $r$=2 indicating that the particles accumulate close to the 3:1 spin-orbit resonance. With increasing grain sizes (panels g, f, and e) the peak becomes more blunt and shifted towards larger distances. Furthermore, a second peak has also been formed at the outer edge of the ring in the case of 5$\mu m$ particles.
Consequently, the smaller the grains size, the narrower the final ring structure.

Since the narrowest ring corresponds to 1$\mu$m grains formed slightly beyond the 3:1 resonance, and it fits best to the proposed place by \citet{Ortiz2017}, the calculations were repeated with longer integration time (T=25000 revs, $\sim$11 years) to see whether the final ring structure is robust enough. In this analysis 60000 non-interacting particles have been involved. Smaller particles, e.g. 0.5 $\mu$ m in size, do not survive the integration at all.

Plotting the number of the particles on bounded orbit versus their orbital period, we obtain the histogram in Figure~\ref{fig:4}. The peaks of the distribution are at certain spin-orbit resonances. The 7:2, 4:1, 9:2 resonances are populated while the 3:1 is almost empty, as expected from escape times in Fig.~\ref{fig:3} panel d. 
The question why then the radial distribution shows a peak at the position of 3:1 spin-orbit resonance (panel b) can be answered by plotting the eccentricity distribution of the ring (Figure~\ref{fig:4} c). The histogram shows that the average eccentricity is 0.12 with 0.065 standard deviation. The initially circular orbits due to the radiation pressure are excited to eccentric motion. It can also been shown that for larger grain sizes the expectation of the eccentricity is smaller. Exactly the same applies when the RP is not taken into account, panel c cyan histogram.

\begin{figure}
	\includegraphics[width=1.0\columnwidth]{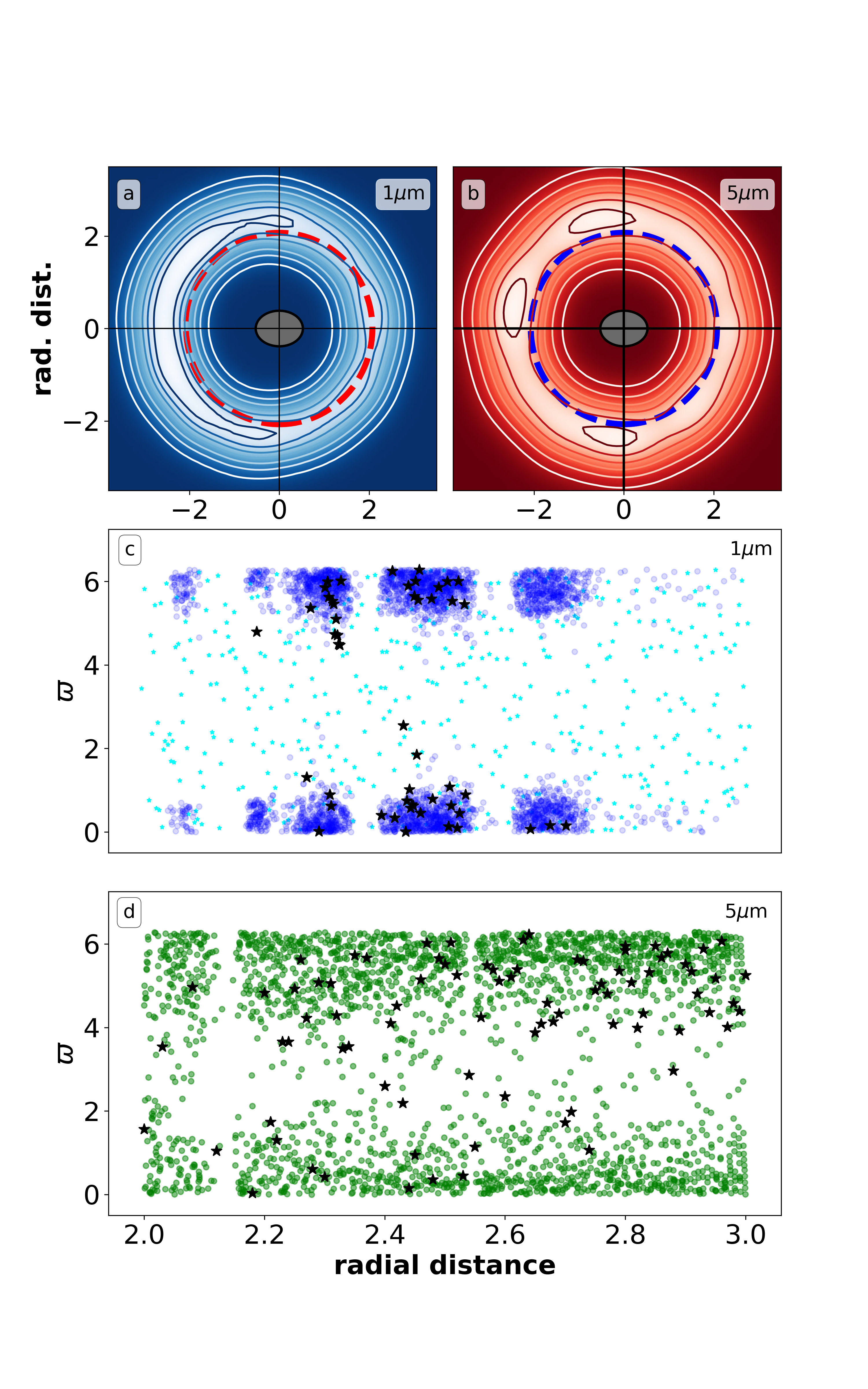}
    \caption{Top: Face-on view of the final stage of the integration. The red dashed line indicates the 3:1 spin-orbit resonance also guiding the eye to eccentric nature of the ring. The gray ellipse represents Haumea to scale. Bottom: The argument of pericentre after the integration (bounded motion only). In the case of 1$\mu$m particles the spin-orbit resonances are well-visible. The accumulation of points around 0$^{\circ}$ suggests an apse-aligned ring. A similar structure has also been found for 5-micron grain size although the azimuthal interval is wider. Black stars are sampling the ring for 1000 year integration. Initial positions of the ring particles: 1.8$\leq r\leq$2.8, 0$\leq \phi\leq$2$\pi.$} % 1687,4566, 5086, 5347 // 314, 1029, 1489, 2396
    \label{fig:5}
\end{figure}

Furthermore, the radial variation of the ring during the transit ingress $\approx$74km and egress $\approx$44km \citep{Ortiz2017} suggests apse-aligned structure that might be caused by self-gravitation or close packing of particles \citep{Goldreich1979,Dermott1980} with some finite value of eccentricity \citep{Nicholson1978}. Here we show that the RP and oblateness can induce the alignment around an oblate body. The precession rate of the orbital ellipses governed by the oblateness of the planet is found to be several hundreds of Haumea's rotation period, $\sim$1-2 months. See for example,  \citet{Kaula1966} (p. 39, Eq.~3.74) or \citet{ssd2000} (p. 497, Eq.~10.45).
%\begin{equation*}
%\frac{\mathrm{d}\varpi}{\mathrm{d}t}=\frac{3nC_{20}R_{\mathrm{e}}^2}{4(1-e^2)^2a^2}[1-5\cos^2i]\sim \mathrm{10^2}\,T_{\mathrm{H}}
%\end{equation*}
%where $\varpi$ is the argument of pericentre, $n$ is the mean motion around the planet (5-20 hours around Haumea) whose mean equatorial radius is $R_{\mathrm{e}}.$ The other parameters $a,e,$ and $i$ are the particle's orbital elements, the semimajor axis, eccentricity, and inclination, respectively. This later is set to be zero. The rate of the pericentre advance becomes several hundreds of Haumea's rotation period, $\sim$1-2 months. 

The first row in Figure~\ref{fig:5} shows the surface density based on the final  configuration of the particles (1 and 5 $\mu$m) after $T$=25000. The different width in radial extent and the eccentricity of the ring is clearly visible in the case of smaller grain size, panel a. 

Panels c and d present the argument of pericentre associated to the panels a and b, respectively. The 1-micron size particles share a pericentre position $\sim$350 degrees in average with 70 degs. standard deviation.  
The blue dots indicate that the longitudes of pericentre are gathering in a narrow band around the longest axis of Haumea. Switching off the radiation pressure this phenomenon ceases and apses cover the whole  2$\pi$ range uniformly (cyan circles). Similar behaviour is depicted in panel d where the pericentres of 5$\mu$m particles group around the same position but the width of the covered domain is wider and apse-alignment is not so strict than in the 1-micron case. Black stars (both panel) indicate 1000 year (ca. 3.5$\times$10$^5$ rotations of Haumea much longer than the precession rate) integration in order to show that the long-lived particles also follow the pericentre conjunction.

We find that both the oblateness of Haumea and RP have significant contribution to the grain dynamics. In brief, the pericentre precession caused by the highly oblate central body protects the very small grains against the strong solar radiation pressure, e.g. \citet{Sfair2009}, and keeps them bounded. And vice-versa, RP excites the eccentricity of individual ring inhabitants helping to align their apses.

\section{Conclusions}

Based on the observations and physical parameters in \citet{Ortiz2017}, the ring dynamics around the dwarf planet Haumea has been investigated numerically. 
In a basic model we take into account a uniformly rotating oblate body whose gravitational potential is described up to second order approximation and the sun's radiation pressure. 

Our results show that different grain sizes produce different ring structures: the smaller the size of the particle, the narrower the ring. We also found that a significant apse-alignment cannot be achieved only by considering the oblateness of the planet. However, if the eccentricity is pumped up to a desired value ($\gtrsim$0.1) the periapse ordering occurs. This job is done by the RP, the stronger the contribution of the RP, the more robust the apse-alignment.  

Moreover, particles in size of 1$\mu$m, initially placed uniformly in wide radial dimension around Haumea, seemingly accumulate circularly near to the 3:1 spin-orbit resonance after surviving a reasonable amount of integration time. We can point out, although it is in good agreement with the proposed radial position of the ring \citep{Ortiz2017}, it is just a coincidence with the maximum accumulation of the particles being on sweeping, individual elliptic orbits trapped into different spin-orbit resonances.

The fact that the escape rate of smaller grains is greater than that of larger ones is caused by dynamical effects including oblateness and radiation pressure. Taking a uniform initial distribution, as we did in our investigation, one can find 70\% deficiency in 1$\mu$m sized particles compared to large 5$\mu$m particles at the end of simulations. The uniform distribution is, however, an unrealistic assumption since several studies (e.g. size distributions of collisional cascades presented in \citet{Dohnanyi1969}) claim a power law distribution (with exponent -3.5)  of particle size, i.e. an excess in $\mu$m size particles. Therefore, it is a reasonable expectation that there are more 1micron-size particles in the ring initially. Based on our results the dominance of smaller grains is plausible as we have seen that only in this case a narrow ring can be formed. Thus, a relationship between the ring geometry and the surviving probability of particles exists for different particle sizes.

The discrepancy of the observed and simulated ring width is two-fold. First, the observed ring profile in visual wavelengths is determined by the optical depth which depends on the properties of the particles (size, density, absorption) and also on surface density. Taking into account this effect one might expect a narrower ring at observations. Moreover, setting the desired optical depth, it is also possible to estimate the ring mass. 

Second, the model we used can be further refined, taking into account escape velocities either at escape or at collision, higher order oblateness parameters ($C_{40},\, C_{42}$), self-gravity, collision, etc., which then might result in more realistic representation of the ring dynamics and characteristics.  

\section*{Acknowledgements}
This work was supported by the NKFIH Hungarian Grants K119993, PD121223 and the Momentum grant of the MTA CSFK Lend\"ulet disc Research Group. TK acknowledges the support of Bolyai Research Fellowship. We also thank the anonymous referee the valuable comments.
%%%%%%%%%%%%%%%%%%%%%%%%%%%%%%%%%%%%%%%%%%%%%%%%%%



\bibliographystyle{mnras}
\bibliography{haumea} % if your bibtex file is called example.bib

%\begin{thebibliography}{99}
%\end{thebibliography}

%%%%%%%%%%%%%%%%%%%%%%%%%%%%%%%%%%%%%%%%%%%%%%%%%%

%%%%%%%%%%%%%%%%% APPENDICES %%%%%%%%%%%%%%%%%%%%%

%\appendix



%%%%%%%%%%%%%%%%%%%%%%%%%%%%%%%%%%%%%%%%%%%%%%%%%%

% Don't change these lines
%\bsp	% typesetting comment
\label{lastpage}
\end{document}